# PGNAA neutron source moderation setup optimization

Zhang Jinzhao[1]（张金钊）Tuo Xianguo[1]（庹先国）

*(1.Chengdu University of Technology Applied Nuclear Techniques in Geoscience Key Laboratory of Sichuan Province，Chengdu 610059，China)*

**Abstract**: Monte Carlo simulations were carried out to design a prompt γ-ray neutron activation analysis (PGNAA) thermal neutron output setup using MCNP5 computer code. In these simulations the moderator materials, reflective materials and structure of the PGNAA $^{252}$Cf neutrons of thermal neutron output setup were optimized. Results of the calcuations revealed that the thin layer paraffin and the thick layer of heavy water moderated effect is best for $^{252}$Cf neutrons spectrum. The new design compared with the conventional neutron source design, the thermal neutron flux and rate were increased by 3.02 times and 3.27 times. Results indicate that the use of this design should increase the neutron flux of prompt gamma-ray neutron activation analysis significantly.

## 1. Introduction

Prompt gamma ray neutron activation analysis (PGNAA) is a rapid, nondestructive, powerful multielemental analysis technique, large samples of some minor, trace light elements and is used in industrial control[1-5]. In a PGNAA analysis, the sample nuclear composition is determined from prompt gamma rays which produced through neutron inelastic scattering and thermal neutron capture. Since the inelastic scattering cross section is small, the PGNAA design based on thermal neutron capture. Low-energy neutrons were absorbed by the target nucleus forming compound nucleus, which has energy equal to the kinetic energy and the neutron capture neutron binding energy. Compound nucleus emits γ photon to go background state in the $10^{-14}$s of time. The gamma yield was determined by the flux of themal neutron, and determined the analysis results. So performing a PGNAA setup depends on thermal neutron flux available at the sample, how to improve the thermal neutron produced ratio of neutron source has been the research focus of PGNAA.

Accelerator neutron, source reactor neutron source, Am-Be neutron source and $^{252}$Cf neutron source can be used as PGNAA neutron source[6]. $^{252}$Cf neutron source which has a high neutron fluxs density and low cost used in PGNAA setup commonly. However, $^{252}$Cf source is isotropic neutron source, to sample less thermal neutron. Neutron source utilization rate is low. In the present study, Monte Carlo calculation was carried out for the design of a $^{252}$Cf neutron source moderation setup for the analysis cement samples[7]. The model of Monte Carlo simulation was verified by experiment[8, 9].We improve the thermal neutron source yield rate of $^{252}$Cf neutron by the PGNAA neutron source structure to the design. The calculation results for the new design were compared with the previous, example: themal neutron flux rate, fast neutron flux rate, gamma rays yeild, results of these studies are presented.

## 2. The model of Monte Carlo simulations

Geometry of the neutrons moderation setup, used in the study, has two forms to compare. Both of the two forms geometry consist of a NaI detector of gamma rays, a cylindrical polyethylene sample compartment and a neutron. Inside diameter of the cylindrical polyethylene sample compartment is 20cm, external diameter is 30cm. Height is 20cm. a neutron source of the previous is only wrapped by a moderator. The new one is different. It has a moderation and a reflector. Reflector places below the neutron source and the a hemispheric moderator are taken on the upper of the neutron source. Our goal: the neutron is reflected back toward the upper portion of the lower part and the utilization ratio of neutrons is improved, because of the different materials for different neutron reflectivity. Fig.1 is a schematic of the previous PGNAA setup source-moderator from paper[7]. Fig.2 is a schematic of new design of this study. We determined the best geometry by calculating.

Supported by National Natural Science Foundation of China(41274109, 41025015)
E-mail:zhangjinzhao_cdut@163.com

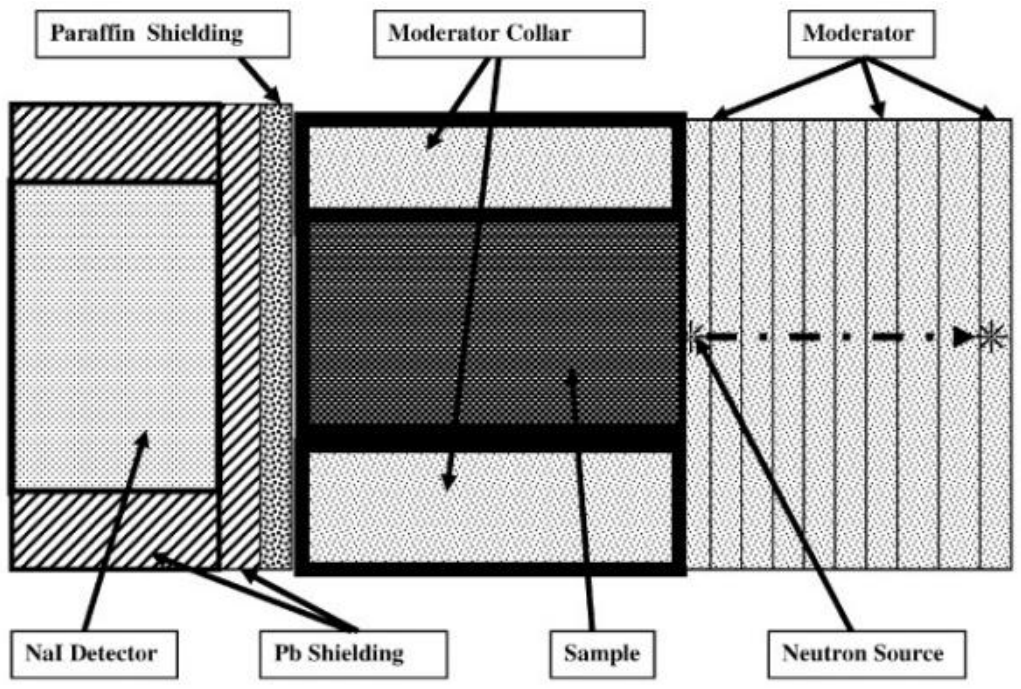

Fig. 1 Schematic of the previous PGNAA setup source-moderator

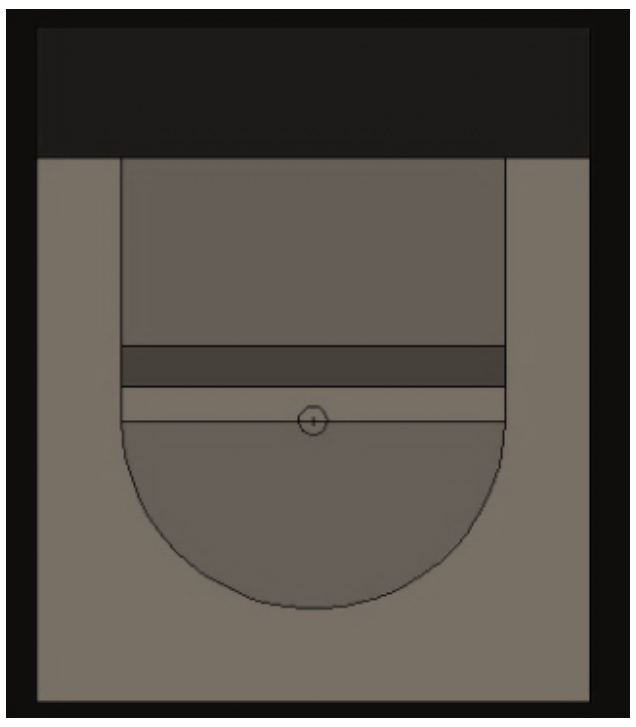

Fig. 2 Schematic of the new PGNAA setup source-moderator of this study

The simulation uses $^{252}Cf$ spontaneous fission neutron source which neutron energy spectrum is a fission spectrum and can be described watt distribution. Formula:

$$f(E) = ce^{-\frac{E}{a}} \sinh\left(\sqrt{bE}\right) a$$

$E$ is the neutron energy, $a = 1.025$, $b = 1.25$, $c = 0.365$. As shown of Figure1, it is a $^{252}Cf$ fission neutron energy spectrum by MCNP5 simulation.

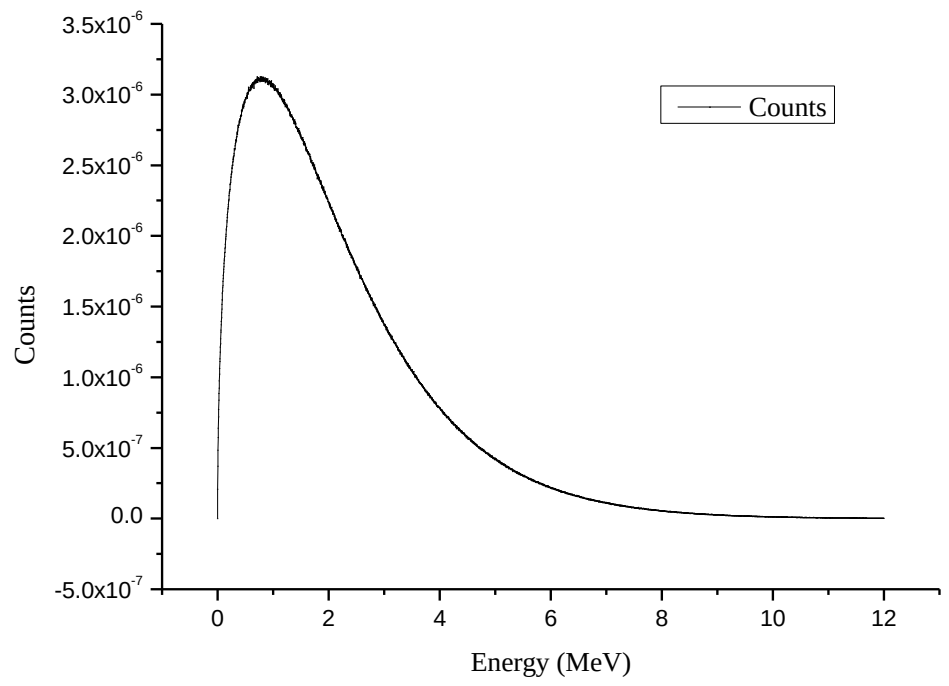

Fig. 3 $^{252}Cf$ Fission neutron spectrum

This study refers to the literature[7, 10] of Portland cement composition as a sample for simulation calculation.

## 2.1 Moderator material Monte Carlo simulation

PGNAA analysis accuracy depends on irradiation neutron fluence rate and neutron capture cross section of to be analyzed nuclides . Radiative capture cross section is the main section for the heavy nuclei and low energy neutron action, the entire cross-section is almost equal with the radiative capture cross section, showing 1/ v . So try to increase the sample thermal neutron fluence rate and reduce the fast neutron fluence rate in the PGNAA analysis. Therefore, we must moderat neutrons, to improve the thermal neutron source neutrons in the proportion, but also pay attention to the material used in moderation to decrease γ background. We chose five materials $H_2O$, $D_2O$, $C_2H_2$, C, Paraffin to calculation. The resluts will be analysised, and the best material will be found out.

Fig.4, Fig.5, Fig.6, Fig.7 is a Monte Carlo calculation reslut of $H_2O$, $D_2O$, $C_2H_2$, C, Paraffin.

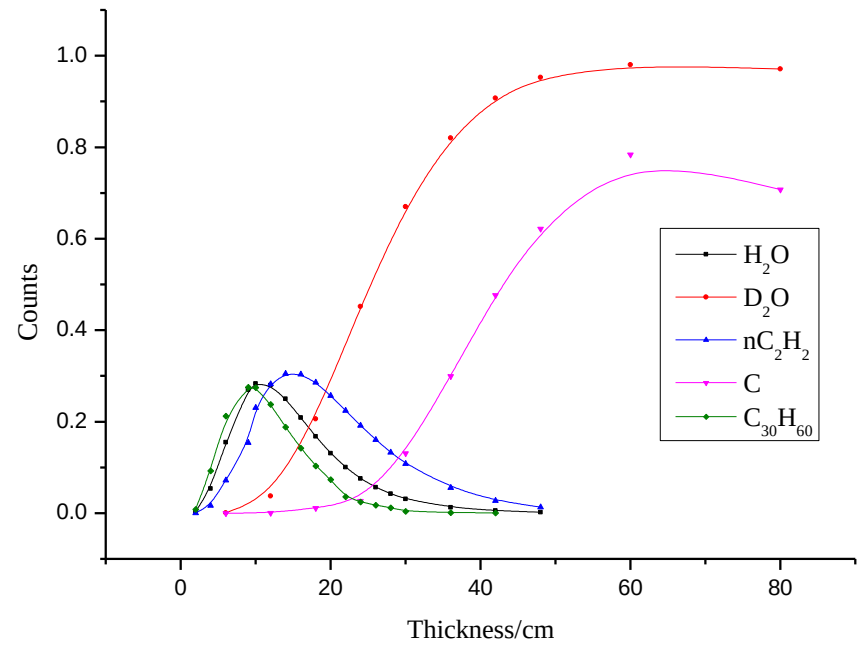

Fig. 4 Thermal neutron flux of five moderator materials increase with the thickness

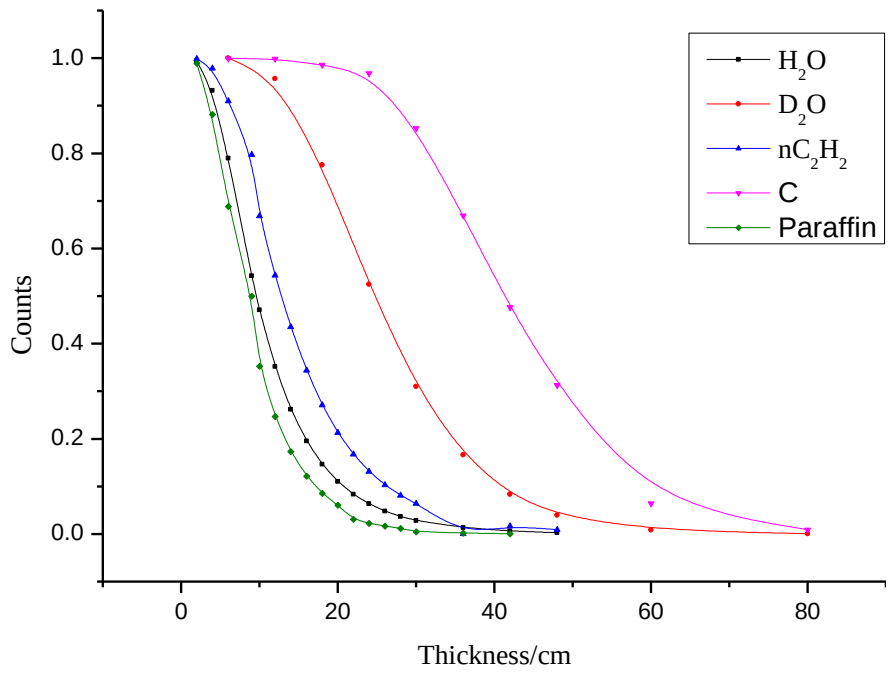

Fig. 5 Fast neutron flux of five moderate materials increase with

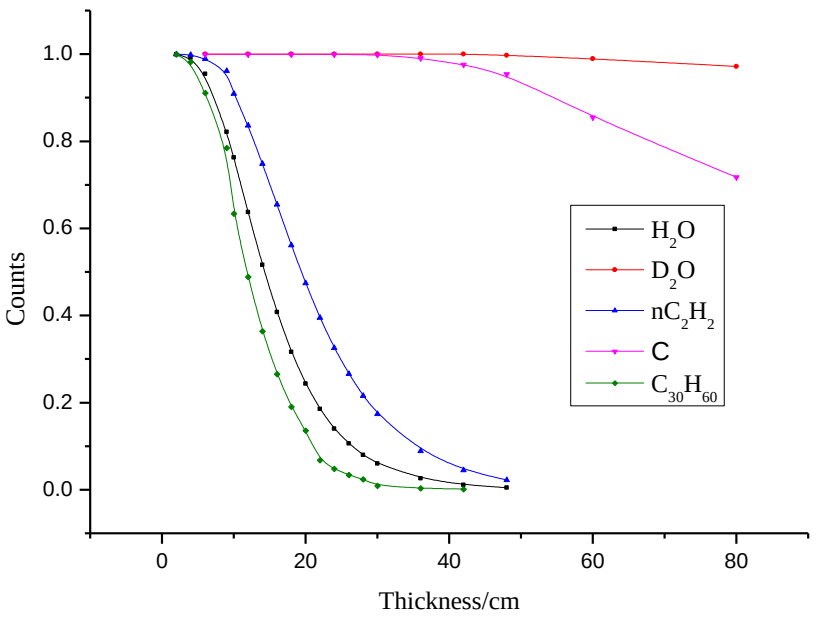

the thickness

Fig. 6 Neutron flux of five moderator materials increase with the thickness

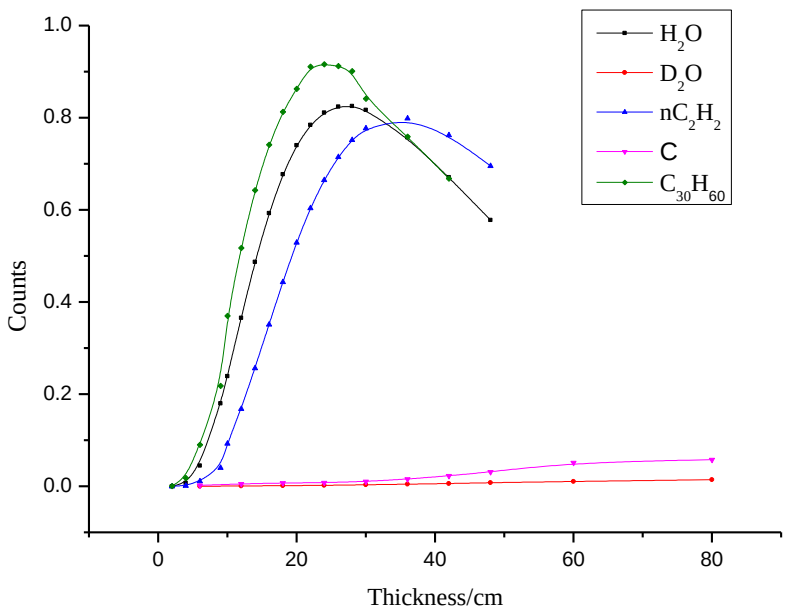

Fig. 7 Gamma ray flux of five moderator materials increase with the thickness

As shown of Figure2, Figure3, Figure4, Figure5:

Thermal neutrons: Thermal neutrons increased rapidly with increasing light water, paraffin, polyethylene thickness in 9cm-12cm growth to the maximum. But it with the moderator increasing the thickness starts to decrease, because a large number of the thermal neutron are absorbed. When heavy water and graphite 48cm, the thermal neutron flux is maximum. Thermal neutrons loss is less with thickness increase in of the moderator.

Fast neutrons: The five substances fast neutron flux is reduced with the increase in thickness of the moderator substance, sequentially: Paraffin, light water, polyethylene, heavy water, graphite.

γ-rays: In a neutron moderator processing, paraffin, light water, polyethylene will produce gamma rays. Yields get maximums and reduce subsequent with the thickness increase in at a thickness of 22cm. Graphite and heavy water is not substantially generated γ-rays.

## 2.2 Reflective material Monte Carlo simulation

The reflection is determined by the scattering cross section of nuclide, the absorption cross section, surface density of atoms and the thickness of the reflector. The higher the scattering cross section is, the smaller absorption cross section is, the higher the surface atom density is, the greater the thickness of the reflector is, the stronger the reflection is. Fig.8, Fig.9 is a Monte Carlo calculation result of $H_2O$, $D_2O$, $C_2H_2$, C, Be, paraffin

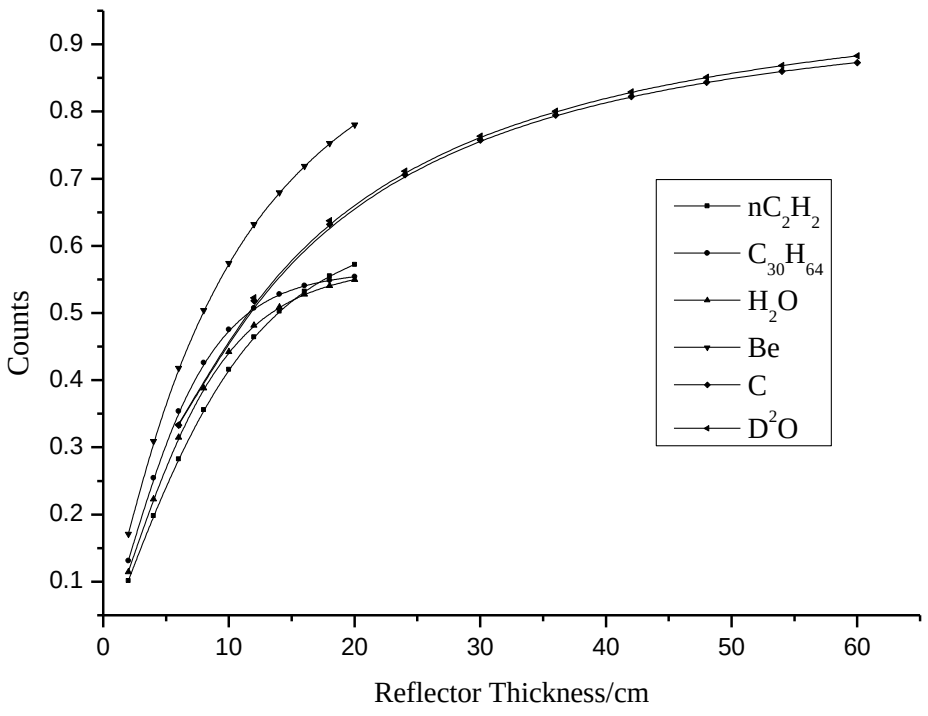

Fig. 8 The reflective neutron flux in different thickness of five kinds of reflective material

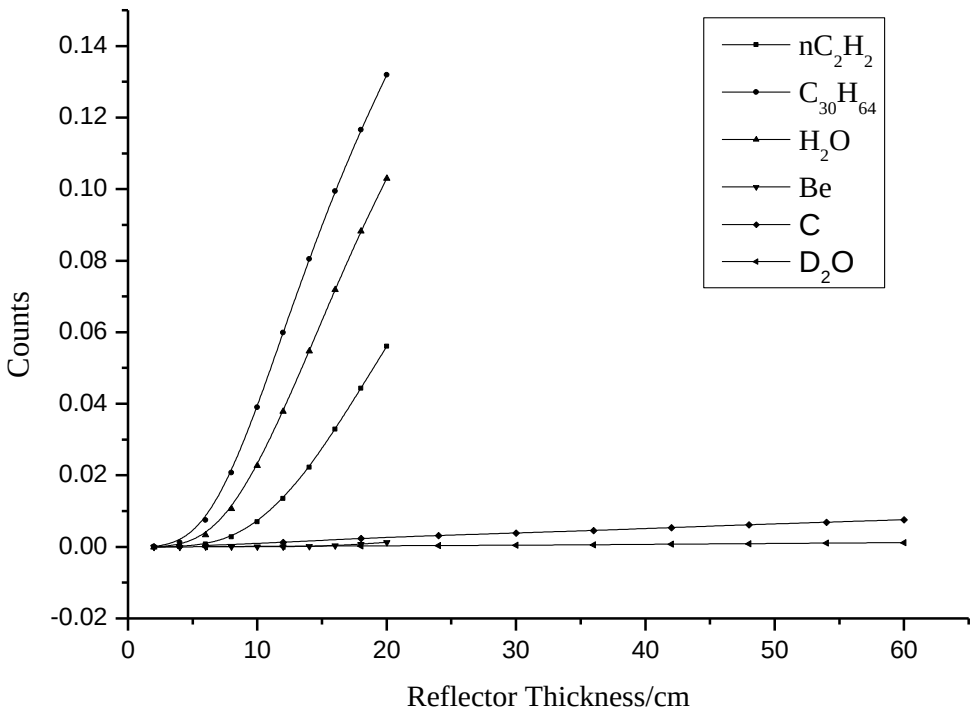

Fig. 9 The γ ray flux in different material and thickness

Fig.8, Fig.9 shows the reflective neutron flux, γ ray in different thickness of five kinds of reflective material. Beryllium is the reflective effect best material on a thin layer, followed by paraffin. Thick layer of heavy water and graphite is reflective of the best materials.The result showed that Thick layer of heavy water and graphite reflection effect was the best. Because it contains hydrogen nuclides, light water, polyethylene, paraffin will absorb neutrons and γ-rays be generated. Beryllium, heavy water and graphite which do not contain neutron absorbing nuclides produces γ-ray flux is extremely low as the reflective materials.

## 3. Neutron source setup optimization

Considering various substances above reflection and moderating effect, design a new neutron source. As thin beryllium reflector best, we use a hemispherical wrapped $^{252}$Cf fission neutron source from the lower part. Thick layer of heavy water and graphite having a similar reflectivity, and they γ-ray yield lower. But heavy water is liquid difficult to perform in practical applications. So we use graphite as a thick layer of the design device reflector. Thick layer of graphite external wrap hemispherical radius is 20cm, so the reflection can be a total reflection.

The upper part of the neutron source was wrapped by Paraffin. Because neutron source is isotropic, so we determined the hemispherical as the geometry of moderation. Paraffin was surrounded by graphite. This design compared with the previous design which used the parallel plate institutions, not only ensured the moderator of the neutron source fully moderated, but also reduced the moderator material for thermal neutron absorption.

The thin beryllium reflector size, which was optimized through thermal neutrons flux rate of the sample. The calculated flux rate of the sample is plotted in Fig.10.

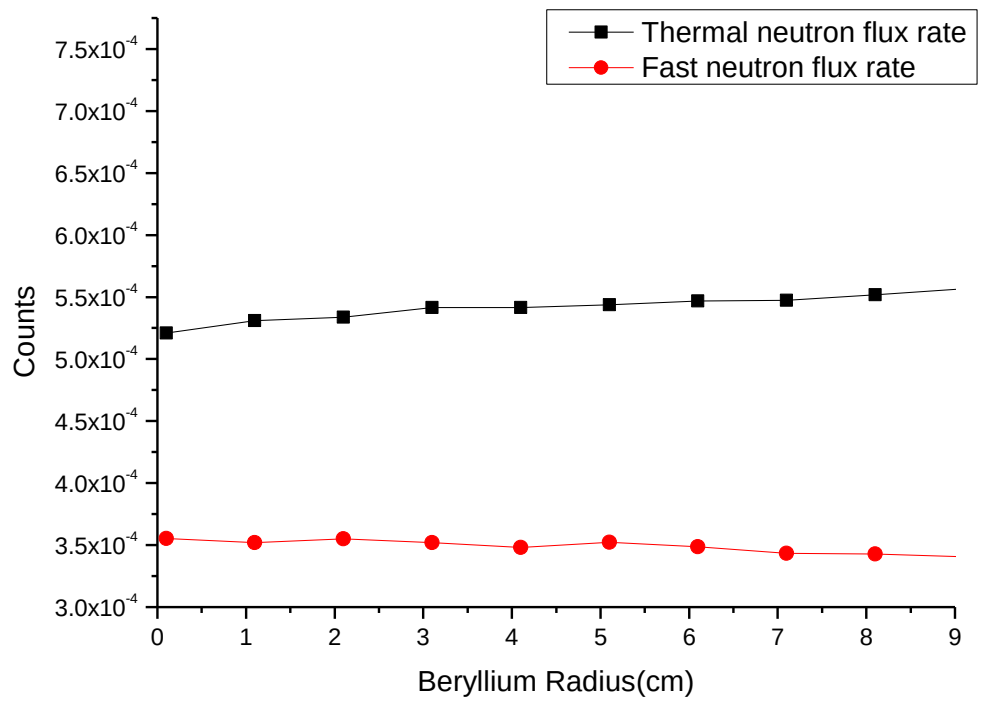

Fig. 10 Thermal neutrons and fast neutrons flux rate of differernt radius beryllium

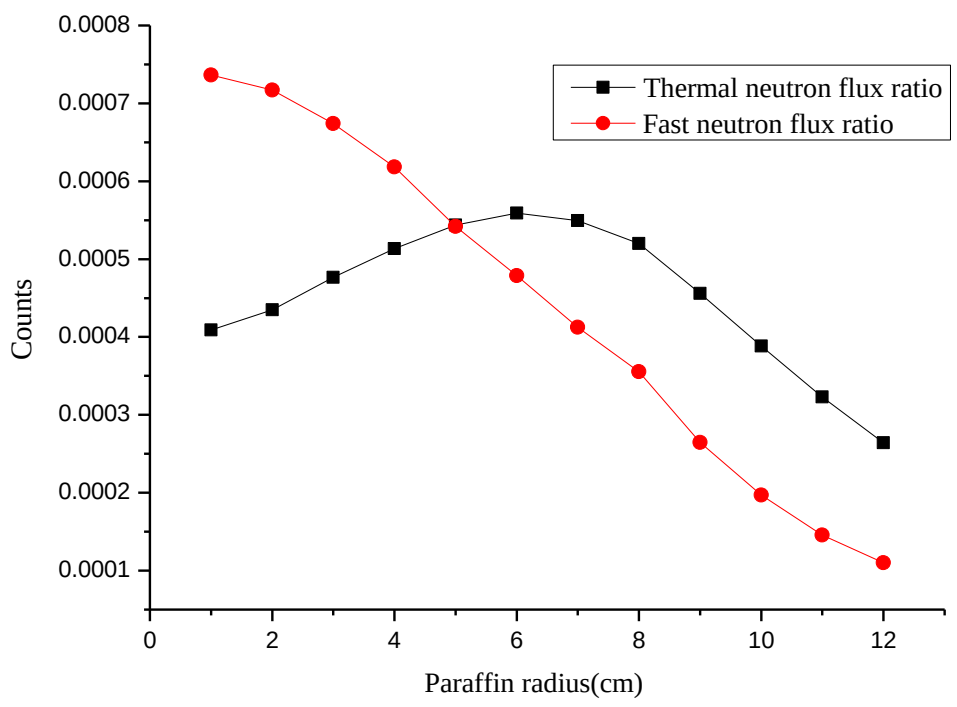

Fig. 11 Thermal neutrons and fast neutrons flux rate of differernt radius of moderation

The results of these calculations are shown in Fig. 10. It is clear that when beryllium thickness increases, the thermal neutron flux increases, fast neutron flux decreases, but the increase and decline are smaller. Due to the price of beryllium is higher, and the effect is not noticeable, so the reflector is all made of graphite.Thermal and fast neutron calculations yield is plotted in Fig.11 as a function of front moderator paraffin radius. When paraffin radius increases, the thermal neutron flux increases, fast neutron flux decreases. The thermal neutron flux rate initially increases with the moderator radius and have a maximum value for 6-7cm radius moderator, then began to decrease. The initial increase in the thermal neutrons yield may be due to neutrons scattering cross section greater than the absorption cross section. Then subsequent decrease in thermal neutron yield with further increase in the moderator radius may be due to the absorption cross section than scattering cross section greater due to increasing thermal neutrons flux rate with increasing moderator radius.The fast neutrons flux rate reduces linearly with the increase of the paraffin radius. As the shown of Fig.11, we can determine, when the paraffin radius of 7cm, reflector of the graphite, the neutron source slowing-down device moderating effect is best.

Tab. 1 The resluts of two designs compares with no sample

| | **Reflection structure** | **Conventional structure** | **Times** |
|---|---|---|---|
| **Thermal flux rate** | 5.49555E-4 | 1.81875E-04 | 3.02E+00 |
| **Fast neutron flux rate** | 4.12318E-4 | 1.26135E-04 | 3.27E+00 |
| **Gamma ray flux rate** | 1.03028E-04 | 3.18849E-05 | 3.23E+00 |
| **The ratio of thermal, fast neutron output flux** | 1.33E+00 | 1.44E+00 | 9.24E-01 |

As shown of Tab.1, in this paper, design of neutron source moderated device compared with Saudi device, the amount of thermal output and fast all have large increase has increased significantly. Thermal neutron output grows by 3.02 times. Fast neutron output grows by 3.27 times. Fast neutron ratio percentage has increased smaller. Purpose is to reduce the proportion of fast neutrons, reducing neutron inelastic collisions nuclides produced by γ-rays, γ to minimize the background. Increase the proportion of thermal neutrons, to improve the thermal neutron capture spectra of the sample measured spectral contribution of the γ, simplified lines, improve the analysis accuracy.the calculated yield of the gamma rays from prompt γ-ray neutron activation is plotted in Fig.12 . As shown of Fig.12, the new structure of the count rate was higher than the previous structure. The energy spectral shape is basically the same. Fig.12 illustrates the new structure to improve the prompt γ count rate which did not make the excitation spectrum become complicated simultaneously. The same neutron source improves the utilization rate of the neutron with the new structure.

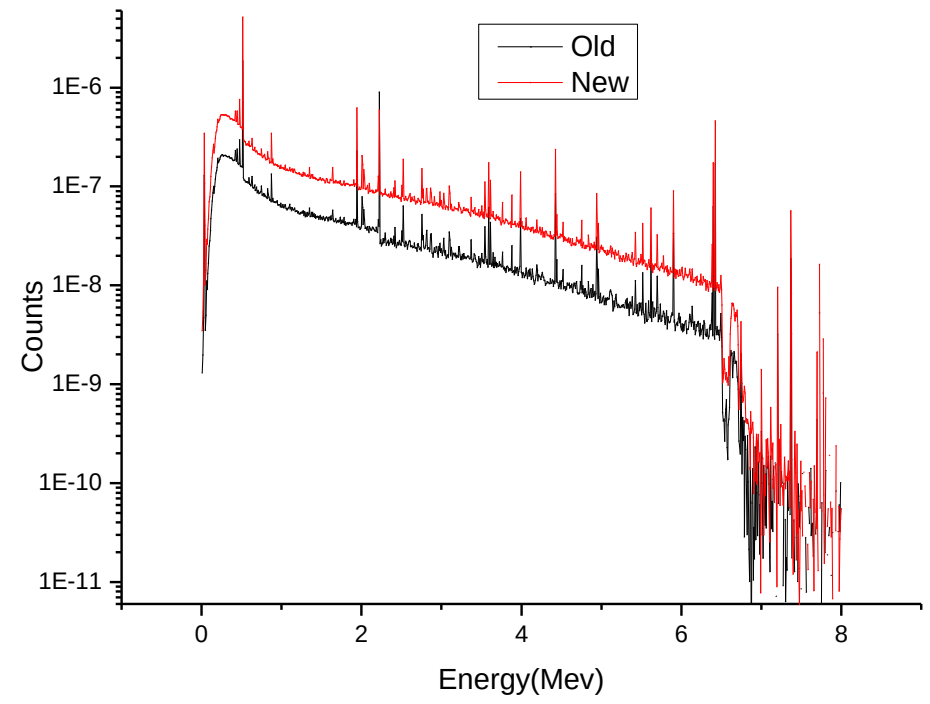

Fig. 12 New and previous design gamma rays yield from prompt γ-ray neutron activation

Fig.13-Fig.22 is the thermal neutron distributions of the two designs in the sample. The samples were divided into 4 × 4 × 4cm grids. The results of these calculations are shown in Fig. 13-Fig.22. It is clear that new design not only on the thermal neutron flux structure has significantly improved, and thermal neutrons in the sample distribution more uniform. It means that the activation of the sample thermal neutrons are more uniform, which is more conducive to large samples and heterogeneous samples for analysis.

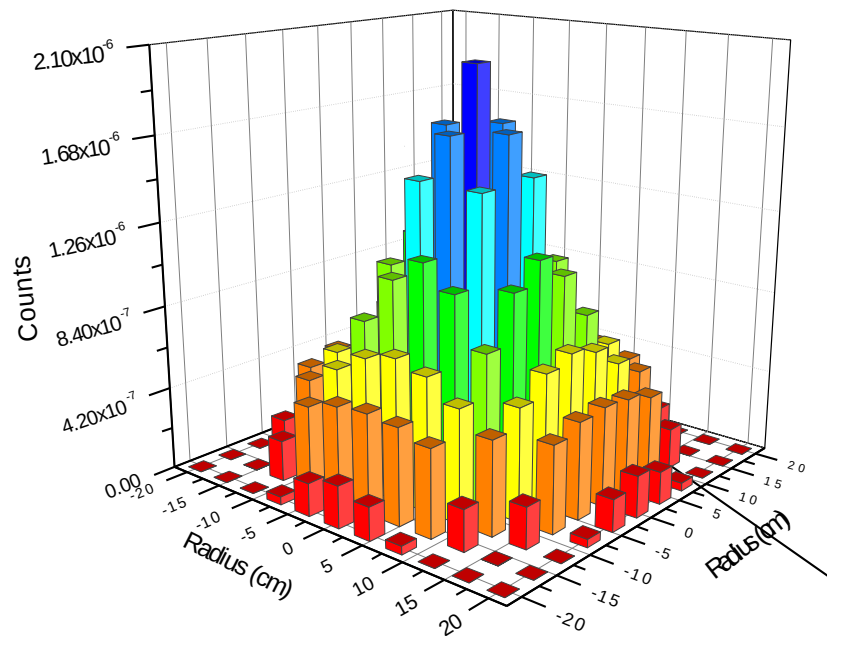

Fig. 13 Previous design thermal neutron flux rate of fist layer

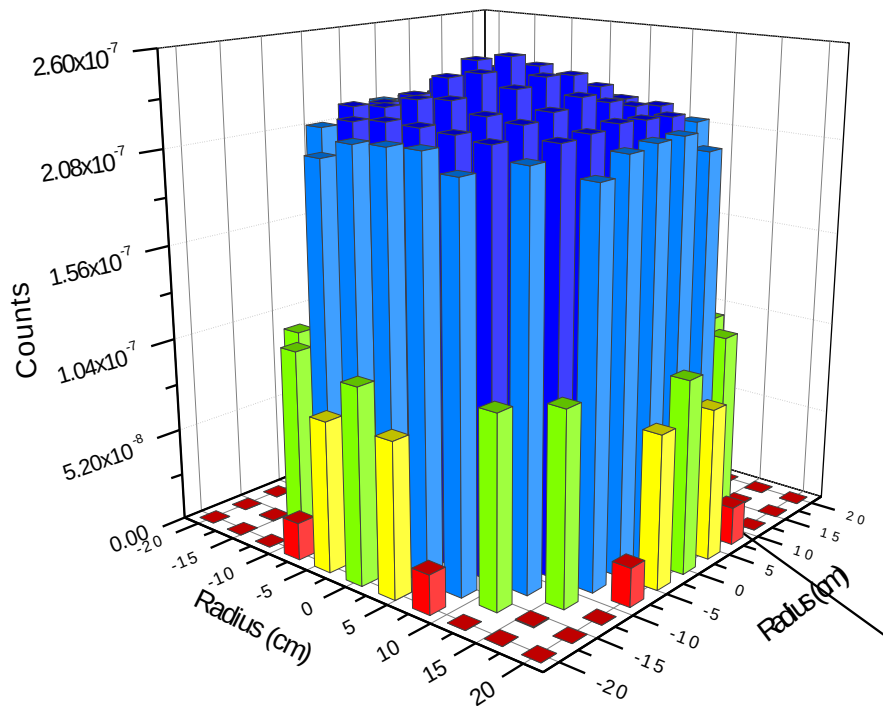

Fig. 17 Previous design thermal neutron flux rate of fifth layer

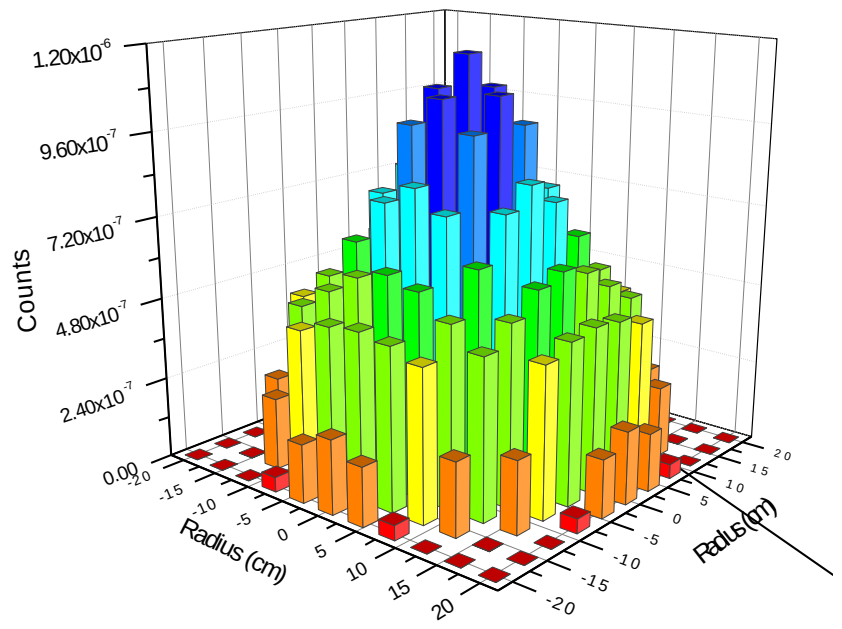

Fig. 14 Previous design thermal neutron flux rate of second layer

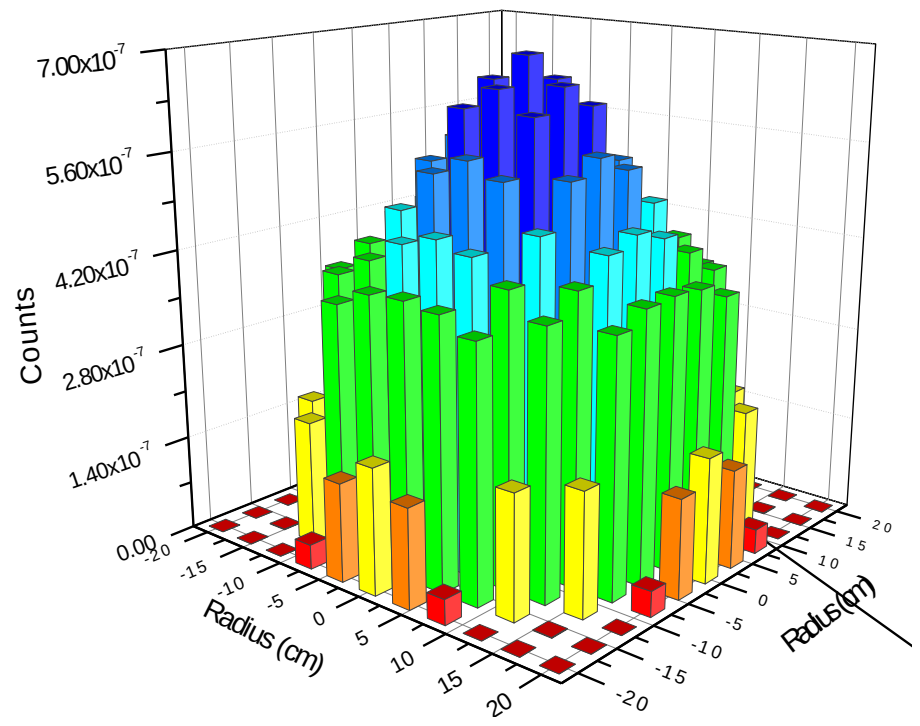

Fig. 15 Previous design thermal neutron flux rate of third layer

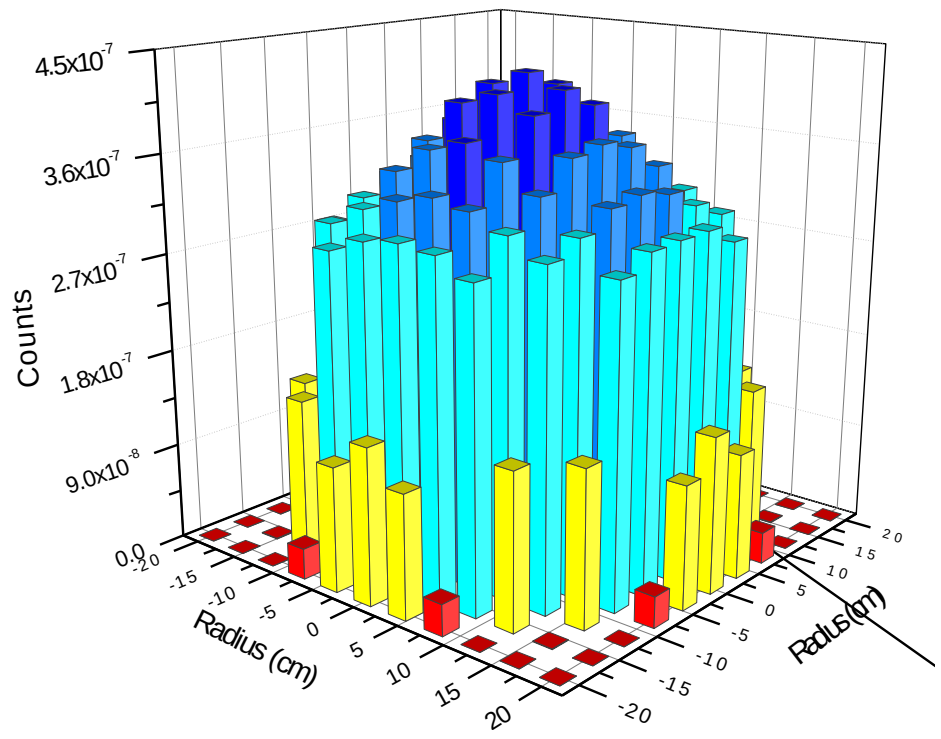

Fig. 16 Previous design thermal neutron flux rate of forth layer

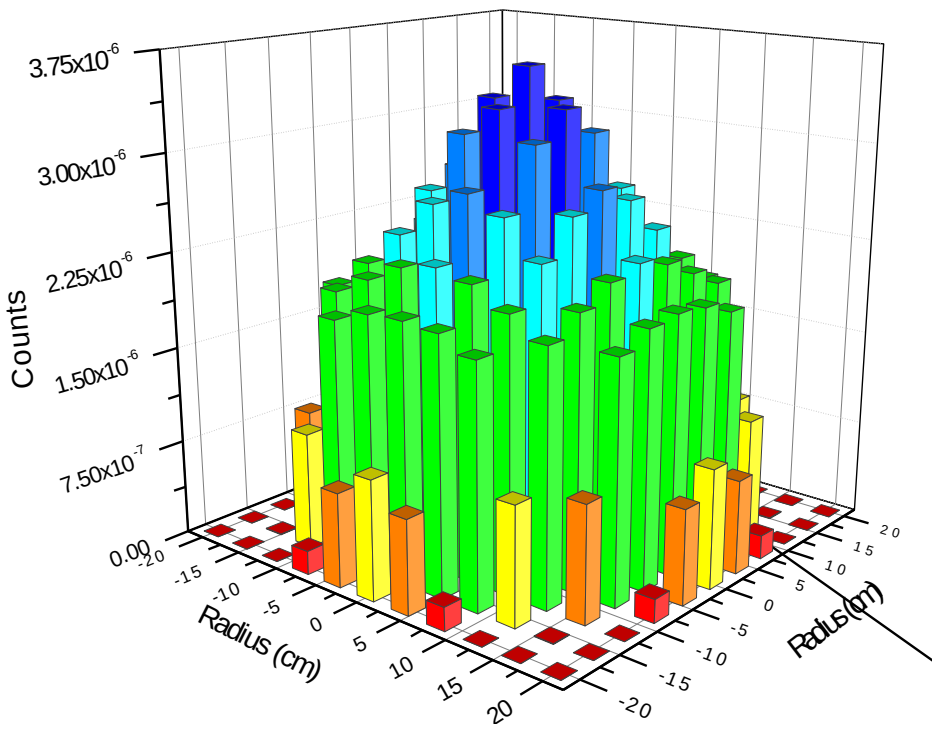

Fig. 18 New design thermal neutron flux rate of first layer

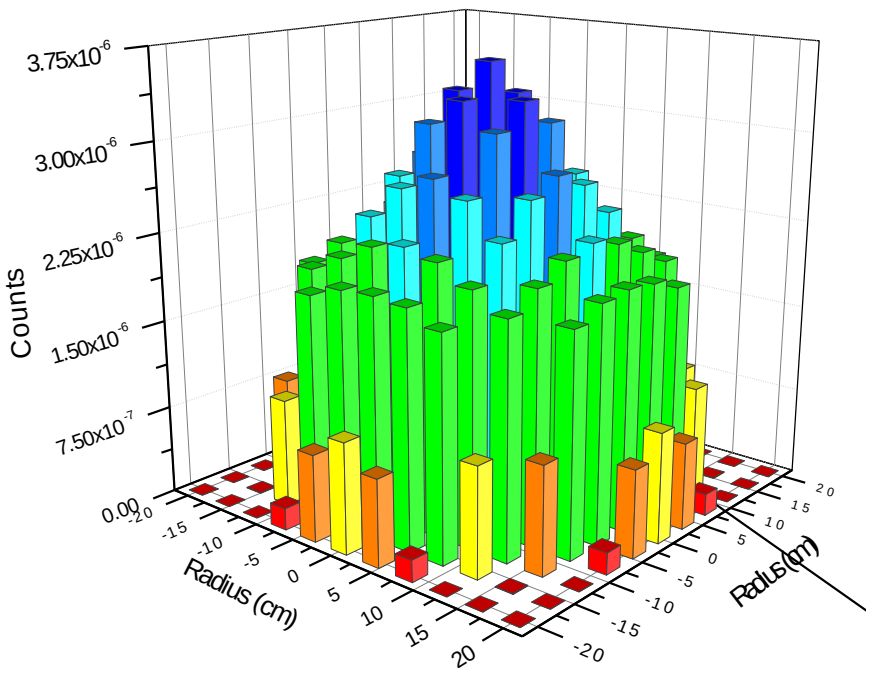

Fig. 19 New design thermal neutron flux rate of second layer

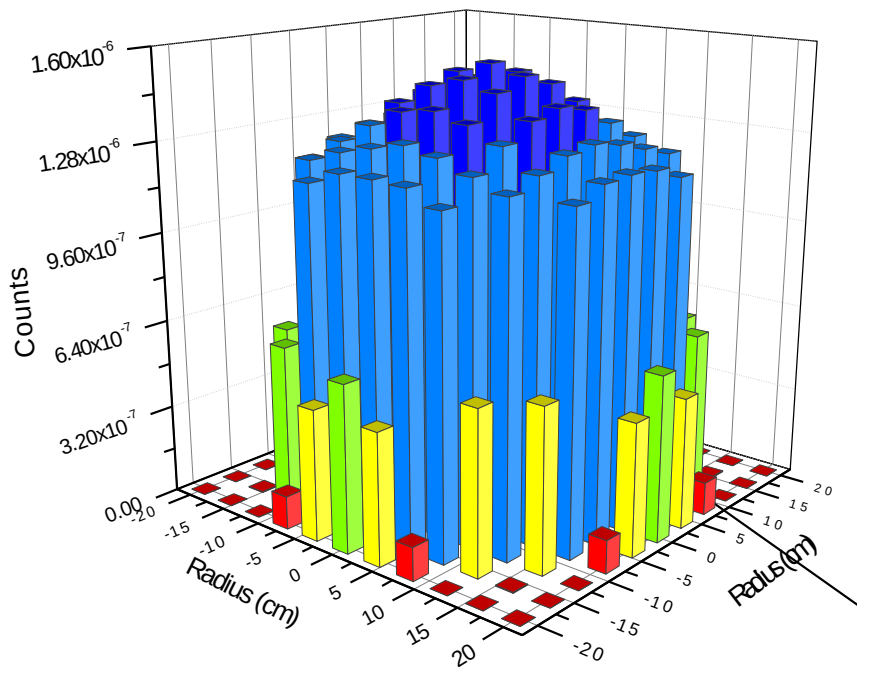

Fig. 20 New design thermal neutron flux rate of third layer

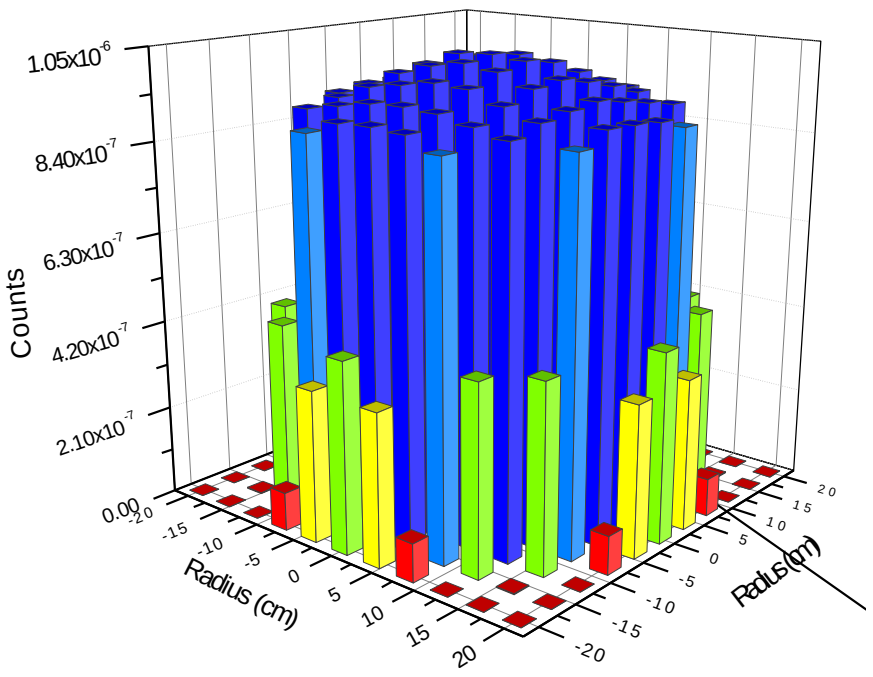

Fig. 21 New design thermal neutron flux rate of forth layer

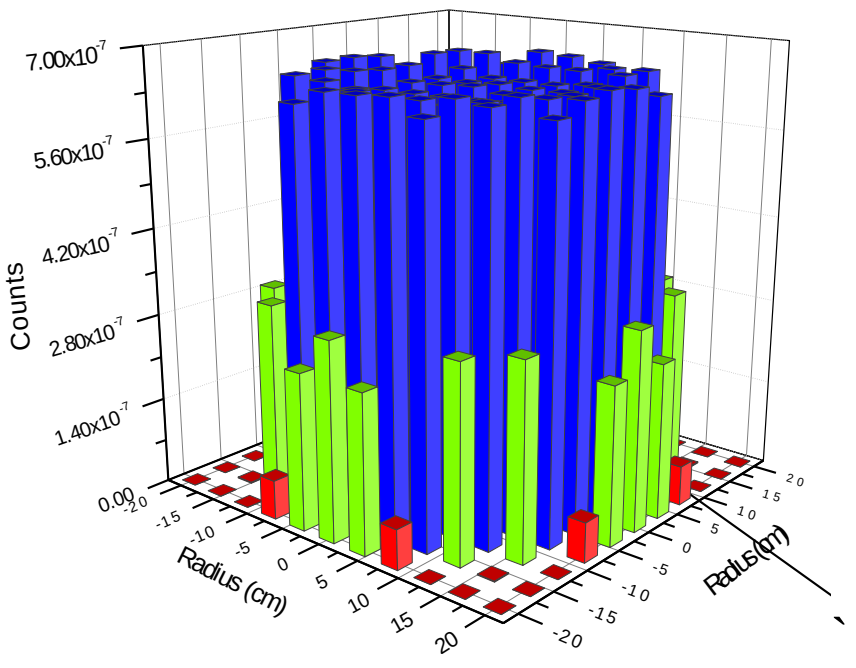

Fig. 22 New design thermal neutron flux rate of fifth layer

## 4. Conclusion

Monte Carlo simulation was carried out to design a PGNAA thermal neutron output setup for increase the thermal neutron output efficiency. The new design PGNAA moderation neutron source device is 3.02 times flux of thermal neutron and 3.27 times flux rate compared with previous device. It improves the utilization of the fission neutron source.

## 5. Acknowledgements

The authors wish to acknowledge the support of the Chengdu University of technology Applied Nuclear Techniques in Geoscience Key and State Key Laboratory of Geohazard Prevention & Geoenvironmental Protection.

# 参考文献